\documentclass[journal=jctcce, manuscript=article, layout=traditional, email=true]{achemso}
\usepackage[T1]{fontenc}
\usepackage{textcomp}
\usepackage{amsmath}
\usepackage{graphicx}
\usepackage{xcolor}
\usepackage[numbers]{natbib}

\makeatletter

\title{Koopmans meets Bethe-Salpeter: Excitonic optical spectra without $GW$.}
\author{Joshua D. Elliott}
\affiliation{Dipartimento di Fisica e Astronomia, Universit\`a Degli Studi di Padova, via Marzolo 8, I-35131 Padova, Italy}
\altaffiliation{CNR-IOM DEMOCRITOS, Consiglio Nazionale delle Ricerche-{}-Istituto
Officina dei Materiali, c/o SISSA, Via Bonomea 265, 34136, Trieste,
Italy}
\email{joshua.elliott@manchester.ac.uk}

\author{Nicola Colonna}
\affiliation{Theory and Simulation of Materials (THEOS) and National Centre for
Computational Design and Discovery of Novel Materials (MARVEL), \'Ecole
Polytechnique F\'ed\'erale de Lausanne, 1015 Lausanne, Switzerland}
\email{nicola.colonna@epfl.ch}

\author{Margherita Marsili}
\affiliation{Laboratoire des Solides Irradi\'es, \'Ecole Polytechnique, CNRS F-91128
Palaiseau, France}

\author{Nicola Marzari}
\affiliation{Theory and Simulation of Materials (THEOS) and National Centre for
Computational Design and Discovery of Novel Materials (MARVEL), \'Ecole
Polytechnique F\'ed\'erale de Lausanne, 1015 Lausanne, Switzerland}

\author{Paolo Umari }
\affiliation{Dipartimento di Fisica e Astronomia, Universit\`a Degli Studi di Padova, via Marzolo 8, I-35131 Padova, Italy}
\altaffiliation{CNR-IOM DEMOCRITOS, Consiglio Nazionale delle Ricerche-{}-Istituto
Officina dei Materiali, c/o SISSA, Via Bonomea 265, 34136, Trieste,
Italy}

\makeatother

\begin{document}

\begin{abstract}
The Bethe-Salpeter Equation (BSE) can be applied to compute from first-principles optical spectra that include the effects of screened electron-hole interactions. 
As input, BSE calculations require single-particle states, quasiparticle energy levels and the screened Coulomb interaction, which are typically obtained with many-body perturbation theory, whose cost limits the scope of possible applications.
This work tries to address this practical limitation, instead deriving spectral energies from Koopmans-compliant functionals and introducing a new methodology for handling the screened Coulomb interaction.
The explicit calculation of the $W$ matrix is bypassed via a direct minimization scheme applied on top of a maximally localised Wannier function basis.
We validate and benchmark this approach by computing the low-lying excited states of the molecules in Thiel's set, and the optical absorption spectrum of a $\text{C}_{60}$ fullerene. 
The results show the same trends as quantum chemical methods and are in excellent agreement with previous simulations carried out at the TD-DFT or $G_{0}W_{0}$-\text{BSE} level. 
Conveniently, the new framework reduces the parameter space controlling the accuracy of the calculation, thereby simplifying the simulation of charge-neutral excitations, offering the potential to expand the applicability of first-principles spectroscopies to larger systems of applied interest. 
\end{abstract}

\section{Introduction}
Charge-neutral excitations play a pivotal role in many environmental and technologically relevant applications, including photovoltaics\cite{kazmerski_photovoltaics:_1997, hagfeldtACR2000} and photocatalysis~\cite{fox_marye_anne_heterogeneous_1993, hagfeldt_light-induced_1995, hoffmann_environmental_1995}. 
Although in principle quantum-mechanical simulations of excited states offer the potential development of novel technologies, the current trade-off between accuracy and computational viability acts as a barrier to the meaningful investigation of realistic systems of interest.

In a simple picture of a charge-neutral excitation, the absorption of a photon promotes an electron to a higher-energy state, creating a hole in the valence manifold. 
It is well known that describing these excitations with density functional theory (DFT) is incorrect.\cite{onida_electronic_2002}
This is due, both to its inability to describe charged excitations (the domain of many-body perturbation theory) and to capture excitonic effects, which are a consequence of two-body interactions between the excited electron-and-hole pair, that renormalize the energy levels and mix the single-particle transitions.

Many-body quantum chemistry methods based on correlated wavefunctions can accurately reproduce experimental data; however, these methods scale poorly with the number of atoms and carry a high computational cost, and viable applications are typically limited to just several tens of atoms.
Besides wavefunction-based methods, two additional {\it ab-initio} approaches are commonly employed in the computation of charge neutral excitations: the time dependent extension of density functional theory~\cite{runge_density-functional_1984,marques_time-dependent_2004} (TD-DFT), and, within the Green's function formalism, the Bethe-Salpeter equation (BSE)~\cite{strinati_application_1986, onida_electronic_2002}. 
TD-DFT and BSE simulations both provide better accuracy-viability trade-offs than wavefunction methods.
In particular, TD-DFT usually performs well for small molecules~\cite{jacqueminJCTC2009}, yet it is known to fail in describing Rydberg states and charge-transfer excitations due to an incorrect description of the asymptotic long range exchange~\cite{tozer_improving_1998,dreuw_failure_2004}.  
Additionally, in solids the long-range exchange should be modulated by the dielectric constant of the system~\cite{bruneval_beyond_2006,botti_time-dependent_2007} making the application of TD-DFT to extended systems more challenging. 
On the other hand, simulations based on the BSE have been shown to reproduce optical properties of molecules~\cite{roccaJCP2010, blase_first-principles_2011, sharifzadeh_quasiparticle_2012, blase_bethesalpeter_2018} and extended systems~\cite{rohlfingPRB2000,onida_electronic_2002}.

\begin{figure}[t]
    \begin{center}
        \includegraphics[]{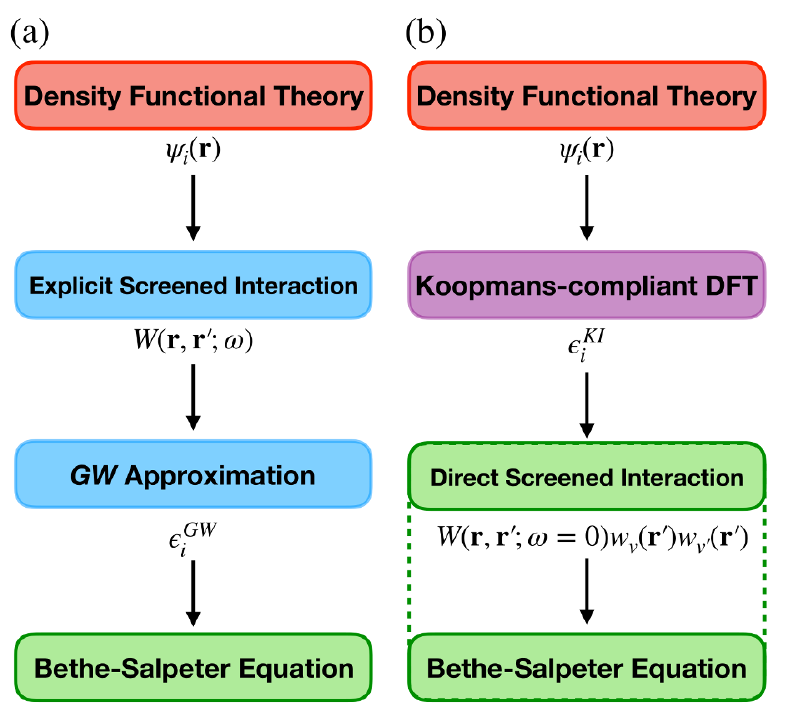}
    \end{center}
    \caption{Schematic representation of the usual and the proposed path to the solutions of the Bethe-Salpeter equation. (a) the most common approach, where eigenvalues and charge screening are derived from the $GW$ approximation (b) the method introduced in this work where eigenvalues are computed via Koopmans-compliant DFT and the action of the screened Coulomb potential is calculated directly via iterative minimization.}
    \label{fig:bse approach}
\end{figure}

Whilst attractive in principle, BSE calculations can be more demanding in practice as they typically involve several steps, each of which can require convergence of several parameters to ensure accuracy in the final result.
For example, the most common approach is based on a combination of DFT and $G_0W_0$ calculations~\cite{hybertsen_electron_1986}, as described in Figure \ref{fig:bse approach}a. 
In the first step, a DFT calculation is used to build a set of single particle Kohn-Sham states, $\psi_i$. 
The second step constructs the frequency-dependent screened Coulomb interaction matrix, $W(\mathbf{r}, \mathbf{r}^\prime ; \omega)$. 
Next comes the quasiparticle equation, which is similar to the DFT Kohn-Sham equations, but includes the self-energy operator in place of the exchange-correlation potential; this is solved for the set of quasiparticle energy levels, $\epsilon_i^{GW}$. 
Only in the final step the BSE is solved for excitonic states, making explicit use of the $\psi_i$'s, $\epsilon_i^{GW}$'s and the screened Coulomb interaction matrix at zero frequency, $W(\mathbf{r}, \mathbf{r}^\prime ; \omega=0)$.

The $GW$ steps (blue in Figure \ref{fig:bse approach}a), which provide $W(\mathbf{r},\mathbf{r}^\prime ;\omega=0)$ and the $\epsilon_i^{GW}$'s, involve extensive and time-consuming convergence testing; in addition they also have an unfavourable $\mathcal{O}(N^4)$ scaling. 
These two factors, combined with the intrinsic complexity of the method, limit the size of system to which the $GW$ and $GW$-BSE methods can be applied. 
Recently, Koopmans-compliant (KC) functionals have emerged as a reliable approach for calculating quasiparticle energies~\cite{daboThesis2008, daboarXiv, daboPRB2010, daboPCCP2013, daboPKN2013, daboBook2014, borghiPRB2014, nguyenPRL2015, nguyenJCTC2016, nguyen_koopmans-compliant_2018, colonna_screening_2018} and as an efficient alternative to more complex electronic structure methods.~\cite{colonna_koopmans-compliant_2019}
Here, we consider bypassing the $GW$ step entirely by following the approach described in Figure \ref{fig:bse approach}b, where the KI functional~\cite{borghiPRB2014, colonna_screening_2018, nguyen_koopmans-compliant_2018, colonna_koopmans-compliant_2019} is used for the computation of quasiparticle energies, $\epsilon_i^\text{KI}$.
For the static screened Coulomb interaction, we eliminate the need to compute the $W(\mathbf{r},\mathbf{r}^\prime ; \omega=0)$ matrix explicitly. 
Instead, we improve on established methods\cite{onida_ab_1995} by introducing a new strategy, which applies the action of the screened 
Coulomb interaction directly onto a basis of Wannier functions, $w_v(\mathbf{r})$, via iterative minimization within the BSE step. 
Not only does this approach remove the need for explicit calculation and storage of the $W$-matrix elements, but it is also parameter free and therefore drastically reduces the computational time required for convergence testing and could expedite future applications in high-throughput materials modelling.

We validate and benchmark this $GW$-free BSE implementation by computing singlet excitations for 
the widely studied Thiel's set~\cite{schreiberJCP2008} and comparing our results with quantum chemical methods, TD-DFT and $GW$-BSE. 
Finally, as an example of larger molecular system we compute the absorption spectrum of C$_{60}$, which is found to be in good agreement with experimental measurement and with previous results from $GW$-BSE. 

\section{Method}

Within many-body perturbation theory, the correlated behavior of electron-hole pairs is described by the Bethe-Salpeter equation: 
The BSE is a Dyson-like equation for the two-particle correlation function~\cite{strinati_application_1986}.
When a static approximation for screening is used, the BSE may be recast in the form of an eigenvalue problem involving a two-body Hamiltonian-like operator $\hat{H}^{eh}$~\cite{rohlfingPRB2000,onida_electronic_2002}. 
Within the Tamm-Dancoff approximation\cite{onida_electronic_2002}, and expressed in the transition space of the electron-hole ($eh$) pair, $\hat{H}^{eh}$ is given by
\begin{equation}
    \hat{H}^{eh}_{vv^\prime cc^\prime} = 
        \hat{D}_{vv^\prime cc^\prime} + 
        \hat{K}_{vv^{\prime}cc^{\prime}}^{x} +  
        \hat{K}_{vv^{\prime}cc^{\prime}}^{d}.
        \label{eq: Heh}
\end{equation}
The indices $v$ \& $v^\prime$ and $c$ \& $c^\prime$ run over occupied and unoccupied single-particle states, and $\hat{D}$, $\hat{K}^x$ and $\hat{K}^d$ are the diagonal, electron-hole exchange and direct screened Coulomb operators that we describe in greater detail below. The eigenvalues and eigenstates of the two-body Schr\"odinger-like equation,
\begin{equation}
    \sum_{v^\prime c^\prime} 
    \hat{H}^{eh}_{vv^\prime cc^\prime} 
    A_{\alpha,v^\prime c^\prime} = 
        E_\alpha\, A_{\alpha,vc},
\end{equation}
represent charge-neutral excitation energies $E_\alpha$ and excitonic (excited) states 
$A_{\alpha,v c}$.
In the language of second quantization a generic excited state in the Tamm-Dancoff approximation may be expressed as a linear
combination of independent particle transitions: 
\begin{equation}
    \vert \Theta \rangle = \sum_{vc} A_{vc}\, \hat{a}_v\hat{a}^\dagger_{c}\vert \Psi_0 \rangle,
\end{equation}
where $\vert \Psi_0 \rangle$ is the ground-state wavefunction, and $\hat{a}^\dagger$ and $\hat{a}$ 
are particle creation and annihilation operators. The diagonal operator of \eqref{eq: Heh} is defined as,
\begin{equation}
    D_{vv',cc'} = 
        (\epsilon_{c} - \epsilon_{v})
        \delta_{vv'}\delta_{cc'},
\end{equation}
where $\epsilon_v$ and $\epsilon_c$ are eigenvalues of the quasiparticle Hamiltonian $\hat{H}^{qp}$.
The excitation energy $E_\alpha$ is renormalized according to the strength of the electron-hole interactions through the electron-hole exchange $\left(K_{vv^{\prime}cc^{\prime}}^{x}\right)$ and direct screened Coulomb $\left(K_{vv^{\prime}cc^{\prime}}^{d}\right)$
potentials~\cite{onida_electronic_2002}. These are respectively,
\begin{equation}
    K_{vv^{\prime}cc^{\prime}}^{x} = 
        \iint\mathrm{d}\mathbf{x}\,\mathrm{d}\mathbf{x}^{\prime}\:
        \psi_{v}^{*}(\mathbf{x})\psi_{c}(\mathbf{x)}\,
        v_{c}(\mathbf{r},\mathbf{r}^{\prime})
        \psi_{v^{\prime}}(\mathbf{x}^{\prime})\psi_{c^{\prime}}^{*}(\mathbf{x}^{\prime})
        \label{eq:electron hole exchange}
\end{equation}
and
\begin{equation}
    K_{vv^{\prime}cc^{\prime}}^{d} = -
        \iint\mathrm{d}\mathbf{x}\,\mathrm{d}\mathbf{x}^{\prime}\:
        \psi_{v}^{*}(\mathbf{x})\psi_{v^{\prime}}(\mathbf{x)}\,
        W(\mathbf{r},\mathbf{r}^{\prime})
        \psi_{c}(\mathbf{x}^{\prime})\psi_{c^{\prime}}^{*}(\mathbf{x}^{\prime}),
        \label{eq:direct screened coulomb}
\end{equation}
where the $\psi_{i}$'s are single particle states, $\mathbf{x}=(\mathbf{r},\sigma)$ is a combined space and spin coordinate, and $v_{c}$ $\left(W\right)$ is the bare (static screened) Coulomb operator.

The solution of the BSE as described is highly demanding; in recent years several methodological advancements have been introduced to lower the computational cost of performing such BSE calculations~\cite{hu_accelerating_2018,roccaJCP2010, marsiliPRB2017}. We will briefly describe our approach, which is implemented in the \textsc{Gwl} code\cite{umariPRB2009,umariPRB2010}, part of the \textsc{Quantum ESPRESSO} distribution~\cite{giannozzi_quantum_2009,giannozzi_advanced_2017}. 
We consider only the case of time-reversal symmetric systems, such that the single-particles states can be taken to be real and a complex-conjugate notation can therefore be neglected. Moreover, we neglect relativistic effects such that $\psi(\mathbf{x})$ factorizes in the product of a spatial function and a spin function and we can retain only the spatial component of the $\mathbf{x}$-coordinate.

Adopting techniques based on density functional perturbation theory~\cite{baroni_phonons_2001} (DFPT), the explicit calculation of unoccupied single-particle states can be removed~\cite{roccaJCP2010,marsiliPRB2017}, and utilizing the so-called
\emph{batch }representation a set of $N_{v}$ single particle functions\citep{roccaJCP2010, walkerPRL2006, marsiliPRB2017}
\begin{equation}
    \xi_{v}(\mathbf{r})=\sum_{c}A_{vc}\psi_{c}(\mathbf{r}),
\end{equation}
are defined for the description of excitonic states such that, in the spatial representation,
\begin{equation}
    \vert \Theta \rangle = \left [ \iint \mathrm{d}\mathbf{r}\, \mathrm{d}\mathbf{r}^\prime \hat{\Psi}(\mathbf{r}) \hat{\Psi}^\dagger(\mathbf{r}^\prime) \Theta(\mathbf{r},\mathbf{r}^\prime) \right ]\vert \Psi_0 \rangle,
\end{equation}
with
\begin{equation}
   \Theta(\mathbf{r},\mathbf{r}^\prime) = \sum_v \psi_v(\mathbf{r})\xi_v(\mathbf{r}^\prime).
\end{equation}
$\hat{\Psi}^\dagger$ and $\hat{\Psi}$ represent the creation and annihilation field operators. In this formulation, the actions of the operators $\left \vert \left \{ \xi_v^\prime \right \} \right \rangle = \hat{D} \left \vert \left \{ \xi_v \right \} \right \rangle $, $\left \vert \left \{ \xi_v^{\prime\prime} \right \} \right \rangle = \hat{K}_{v,v^\prime}^x \left \vert \left \{ \xi_v \right \} \right \rangle$ and $\left \vert \left \{ \xi_v^{\prime\prime\prime} \right \} \right \rangle = \hat{K}_{v,v^\prime}^d \left \vert \left \{ \xi_v \right \} \right \rangle $ can be written, elementwise, as~\cite{marsiliPRB2017}
\begin{eqnarray}
    \xi_v^\prime(\mathbf{r}) &=& 
        \left (
        \hat{H}^{qp} - \epsilon_v I
        \right )
        \xi_v(\mathbf{r}), 
        \label{eq: Diagonal}\\
    \xi_{v}^{\prime\prime}(\mathbf{r}) &=&
        \int\mathrm{d}\mathbf{r}^{\prime}\,
        P_{c}(\mathbf{r},\mathbf{r}^{\prime})\psi_{v}(\mathbf{r}^{\prime})
        \sum_{v^\prime} \left [ 
        \int \mathrm{d}\mathbf{r}^{\prime\prime}\,
        v_{c}(\mathbf{r}^{\prime},\mathbf{r}^{\prime\prime})
        \psi_{v^{\prime}}(\mathbf{r}^{\prime\prime})
        \xi_{v^{\prime}}(\mathbf{r}^{\prime\prime}) \right ],\\
    \xi_{v}^{\prime\prime\prime}(\mathbf{r}) &=&
        -\int\mathrm{d}\mathbf{r}^{\prime}\,
        P_{c}(\mathbf{r},\mathbf{r}^{\prime})
        \sum_{v^{\prime}}\xi_{v^{\prime}}(\mathbf{r}^{\prime})\,
        \left[
        \int\mathrm{d}\mathbf{r}^{\prime\prime}\,
        W(\mathbf{r}^{\prime},\mathbf{r}^{\prime\prime})
        \psi_{v^{\prime}}(\mathbf{r}^{\prime\prime})\psi_{v}(\mathbf{r}^{\prime\prime})
        \right],\label{eq:wannier phi}
\end{eqnarray}
where we introduce the operator $P_c$ for projections onto the manifold of empty states,
\begin{equation}
    P_{c}(\mathbf{r},\mathbf{r}^{\prime})=
        \sum_{c}\psi_{c}(\mathbf{r})\psi_{c}(\mathbf{r}^{\prime})=
            \delta(\mathbf{r}-\mathbf{r}^{\prime})-\sum_{v}\psi_{v}(\mathbf{r})\psi_{v}(\mathbf{r}^{\prime}).
\end{equation}
The advantage of this approach is that explicit sums over the empty-state manifold, which enter into the evaluation of $W$ and constitute a significant computational bottleneck, are removed~\cite{roccaJCP2010}.

A further speed-up (and improvement to overall scaling\cite{marsiliPRB2017}) can be obtained by noting that, unlike $\xi_v^{\prime\prime}$ that is evaluated in reciprocal space, it is more convenient to compute $\xi_v^{\prime\prime\prime}$ in real space, using a maximally localised Wannier Function (MLWF) representation for the occupied single-particle states. A given valence state is transformed to a MLWF via a unitary rotation~\cite{marzariPRB1997},
\begin{equation}
    w_{v}(\mathbf{r})=\sum_{v^{\prime}}U_{vv^{\prime}}\psi_{v^{\prime}}(\mathbf{r}),
\end{equation}
by the matrix $U$. In the present implementation we use the algorithm proposed by Gygi et al.\citep{gygiCPC2003} 
The batches, $\xi_v$, are also reversibly transformable in the MLWF representation:
\begin{equation}
    \tilde{\xi}_{v}(\mathbf{r}) = 
        \sum_{v^{\prime}}U_{vv^{\prime}}\xi_{v}(\mathbf{r}).
\end{equation}
Thus, we can rewrite Eqn. \eqref{eq:wannier phi} as
\begin{equation}
    \tilde{\xi}_{v}^{\prime\prime\prime}(\mathbf{r})=
        -\int\mathrm{d}\mathbf{r}^{\prime}\,
        P_{c}(\mathbf{r},\mathbf{r}^{\prime})
        \sum_{v^{\prime}}\tilde{\xi}_{v^{\prime}}(\mathbf{r}^{\prime})\,
        \left[\int\mathrm{d}\mathbf{r}^{\prime\prime}\,
        W(\mathbf{r}^{\prime},\mathbf{r}^{\prime\prime})
        w_{v^{\prime}}(\mathbf{r}^{\prime\prime})w_{v}(\mathbf{r}^{\prime\prime})
        \right].\label{eq:wannier W}
\end{equation}
Exploiting locality, we define a threshold for which a given pair of MLWFs overlap,
\begin{equation}
    s < 
        \int\mathrm{d}\mathbf{r}\,
        \vert w_{v}(\mathbf{r})\vert^{2}\vert w_{v^{\prime}}(\mathbf{r})\vert^{2}.
    \label{eq: s threshold}
\end{equation}
By excluding non-overlapping pairs of MLWFs from the summation in Eqn. (\ref{eq:wannier W}),
it becomes possible to lower the scaling of the evaluation of $\xi_v^{\prime\prime\prime}$ from $\mathcal{O}(N^{4})$
to $\mathcal{O}(N^{3})$~\cite{marsiliPRB2017}.

A given BSE calculation therefore requires as input the zero-frequency screened Coulomb interaction (Equation \ref{eq:wannier W}), quasiparticle eigenvalues, (Equation \ref{eq: Diagonal}) and a set of single-particle states. 
The screened interaction and quasiparticle energies for valence and conduction states could be provided for example from a \textsc{Gwl} $G_0W_0$ calculation, while the single-particle states are eigenstates of the Kohn-Sham Hamiltonian computed using the \textsc{pw.x} code of \textsc{Quantum ESPRESSO}. 
The relatively long computation times and extensive convergence testing required make the $GW$-steps in Figure \ref{fig:bse approach}a a bottleneck for BSE calculations.
In the next sections, we show how the use of the $GW$-approximation can be bypassed, first, by coupling the BSE with KC functionals, and then introducing an iterative minimization scheme to calculate  the action of the screened Coulomb interaction on the basis of MLWFs.

\subsection{Quasiparticle energies from Koopmans-compliant functionals}

In recent years, KC functionals have emerged as a reliable and efficient alternative to Green's function methods for the prediction of photoemission properties both in finite~\cite{borghiPRB2014,nguyenPRL2015,nguyenJCTC2016,colonna_screening_2018,colonna_koopmans-compliant_2019} and extended systems~\cite{nguyen_koopmans-compliant_2018}.
In this section, we briefly review the key features of the KC functionals; the interested reader is referred to previous work~\cite{daboPRB2010,borghiPRB2014,borghi_variational_2015} for a detailed description of the method.

In a nutshell, any approximate DFT functional is made Koopmans-compliant by enforcing a generalized condition of piecewise linearity (PWL) to the entire electronic manifold.
This is achieved by removing, orbital-by-orbital, the non-linear dependence of the total energy as a function of fractional occupations, and replacing it with a linear Koopmans' term.
There are several different definitions available for the slope of the linear term; in this work we focus on the KI\citep{borghiPRB2014} functional ( ``I'' standing for ``integral''), for which the slope is chosen as the difference between the energies of the two adjacent electronic configurations with integer occupation. 
This translates the $\Delta$ self-consistent-field ($\Delta$SCF) approach to a functional form. Starting from an approximate DFT functional $E^{{\rm DFT}}[\rho]$, the KI functional is obtained as 
\begin{equation}
    E^{{\rm KI}} = 
        E^{{\rm DFT}}+\sum_{i}\alpha_{i}\Pi_{i}^{{\rm KI}},
    \label{eq: E_KI}
\end{equation}
with 
\begin{eqnarray}
    \Pi_{i}^{{\rm KI}} &=& 
        -\int_{0}^{f_{i}}ds
            \langle\psi_{i}|H^{{\rm DFT}}(s)|\psi_{i}\rangle + 
        f_{i}\int_{0}^{1}ds
            \langle\psi_{i}|H^{{\rm DFT}}(s)|\psi_{i}\rangle, \nonumber \\
    &=& 
        -\{E_{{\rm Hxc}}^{{\rm DFT}}[\rho] - 
        E_{{\rm Hxc}}^{{\rm DFT}}[\rho-\rho_{i}]\} + 
        f_{i}\left\{E_{{\rm Hxc}}^{{\rm DFT}}[\rho-\rho_{i}+n_{i}] - 
        E_{{\rm Hxc}}^{{\rm DFT}}[\rho-\rho_{i}]\right\}.
    \label{eq: bare koopman potential}
\end{eqnarray}
Here, $\rho_{i}=f_{i}|\psi_{i}(\mathbf{r})|^{2}=f_{i}n_{i}(\mathbf{r})$, is the density of the $i^\text{th}$ orbital, $H^{{\rm DFT}}$ is the Kohn-Sham Hamiltonian associated to the underlying DFT functional, and $E_{{\rm Hxc}}^{{\rm DFT}}[\rho]$ is the Hartree exchange and correlation energy of the approximate functional at density $\rho$. 
The bare $\Pi_{i}^{{\rm KI}}$ correction of equation \eqref{eq: E_KI} imposes a generalized PWL in a frozen orbital picture, i.e. assuming that the orbitals do not readjust when the occupation of one of these is changed.
Relaxation effects associated with the addition/removal process are accounted for by the screening coefficients $\alpha_{i}$ renormalizing the bare Koopmans corrections.

The direct variation of the Koopmans' contribution to the total energy with respect to the each orbital density leads to orbital-dependent potentials, 
\begin{equation}
    v_{i}^{{\rm KI}} = 
        \alpha_{i}\frac{\delta\Pi^{\rm KI}_{i}}{\delta\rho_{i}(\mathbf{r})},
\end{equation} 
that can be interpreted as quasiparticle approximations to the spectral potential~\cite{ferretti_bridging_2014, colonna_koopmans-compliant_2019}, i.e. the local, frequency-dependent potential necessary and sufficient to describe spectral properties~\cite{gatti_transforming_2007,vanzini_spectroscopy_2018}. 

Similarly to the case of a $G_0W_0$ calculation, starting from the single-particle orbitals and energies of the underlying density functional, the KI energies are computed as, 
\begin{equation}
    \varepsilon_{i}^{{\rm KI}} = 
        \varepsilon_{i}^{{\rm DFT}} + 
        \langle\psi_{i}^{{\rm DFT}}|v_{i}^{{\rm KI}} | \psi_{i}^{{\rm DFT}}\rangle = 
        \varepsilon_{i}^{{\rm DFT}}+\Delta\varepsilon_{i}^{{\rm KI}}. 
    \label{eq: KI eigenvalues}
\end{equation}
The additional computational costs associated with the evaluation of Koopmans' terms $\Delta\varepsilon_{i}^{{\rm KI}}$ are mainly due to the calculation of the screening coefficients $\alpha_i$. This is because, as shown in Eqn.~\eqref{eq: bare koopman potential}, the bare Koopmans' potential is expressed in terms of the Hxc energies
and potentials of the underlying functional at densities $\rho(\mathbf{r})$
and $\rho(\mathbf{r})\,\pm \,n_{i}(\mathbf{r})$ (depending whether a
particle is removed ($-$) or added ($+$) to the system); these quantities are readily available from the ground-state calculation.

The screening coefficients are given by~\cite{colonna_screening_2018}
\begin{equation}
    \alpha_i = 
        \frac{\langle n_i | \epsilon_{\rm DFT}^{-1}f_{\rm Hxc} | n_i \rangle}{\langle n_i | f_{\rm Hxc} | n_i \rangle},
    \label{eq.alpha}
\end{equation}
where $\epsilon_{\rm DFT}(\mathbf{r},\mathbf{r}')$ is the static dielectric matrix evaluated at the DFT level and $f_{\rm Hxc}(\mathbf{r},\mathbf{r}')=\frac{\delta^2 E_{\rm Hxc}^{\rm DFT}} {\delta \rho(\mathbf{r})\delta\rho(\mathbf{r}')}$ is the static Hartree-exchange-and-correlation kernel.
In principle the evaluation of each screening coefficient requires the solution of a static linear-response problem~\cite{colonna_screening_2018}, and can be efficiently computed resorting to the machinery of DFPT~\cite{baroni_phonons_2001}. 
A set of self consistent equations analogous to the KS ones have to be solved for each $\alpha_i$; the resulting computational cost of the approach scales as $\mathcal{O}(N^4)$. 
In practice, several heuristic recipes can be derived to predict those; 
although not pursued in this work, the number 
of linear-response calculations can be greatly reduced
by exploiting the symmetry of the systems and the fact that there
exists a strong connection between screening coefficients and the
degree of localization of the orbitals: orbitals with similar
spatial extension have very similar screening coefficient~\cite{colonna_screening_2018,nguyen_koopmans-compliant_2018}.

\subsection{Direct calculation of static screened Coulomb interaction}
\label{sec: screening}

In the present work, we propose a method that enables fast calculations
of the fully converged static screened Coulomb interaction based on a direct minimization scheme.
The static screened Coulomb interaction $W\left(\mathbf{r},\mathbf{r}^{\prime}\right)$ appears in the evaluation of the direct operator $K^{d}$; specifically in the squared bracketed term in Eqn. (\ref{eq:wannier W}). In our
previous approach~\cite{marsiliPRB2017}, and in the literature, the
treatment of $K^{d}$ is carried out in two independent steps according
to the separation of $W(\mathbf{r},\mathbf{r}^{\prime})$
into $(i)$ evaluation of the bare interaction operator, $v_{c}(\mathbf{r},\mathbf{r}^{\prime})$,
and $(ii)$ the electron correlation $W_{c}(\mathbf{r},\mathbf{r}^{\prime})$:
\begin{equation}
    W(\mathbf{r},\mathbf{r}^{\prime}) =
        v_{c}(\mathbf{r},\mathbf{r}^{\prime}) + 
        W_{c}(\mathbf{r},\mathbf{r}^{\prime}). 
    \label{eq:W}
\end{equation}
First, we examine the electron
correlation, which can be expressed in terms of the reducible screened
polarizability, $\Pi\left(\mathbf{r},\mathbf{r}^{\prime}\right)$.
Introducing a dot to denote the product of operators $v\cdot\Pi=\int\mathrm{d}\mathbf{r}^{\prime\prime}v\left(\mathbf{r},\mathbf{r}^{\prime\prime}\right)\Pi\left(\mathbf{r}^{\prime\prime},\mathbf{r}^{\prime}\right)$,
we write
\begin{equation}
    W_{c} = 
        v_{c}\cdot\Pi\cdot v_{c}.
    \label{eq:Wc}
\end{equation}
The Dyson equation
\begin{equation}
    \Pi = 
        P + P\cdot v_{c}\cdot\Pi
\end{equation}
defines $\Pi\left(\mathbf{r},\mathbf{r}^{\prime}\right)$ in terms
of the irreducible polarizability $P\left(\mathbf{r},\mathbf{r}^{\prime}\right)$,
which is usually calculated in the random-phase approximation as the
product of two single particle propagators.\citep{marsiliPRB2017,umariPSSB2011}
By stating explicitly the terms that constitute the infinite summation
of $\Pi\left(\mathbf{r},\mathbf{r}^{\prime}\right)$:
\begin{equation}
    \Pi = 
        P + P\cdot v_{c}\cdot P + 
        v_{c}\cdot P\cdot v_{c}\cdot P\cdot v_{c}\dots,
    \label{eq:reducible screened polarizability}
\end{equation}
and then inserting Eqns. (\ref{eq:Wc}) and (\ref{eq:reducible screened polarizability})
into the expression for the screened Coulomb potential defined in
Eqn. (\ref{eq:W}), it can be seen that $W$ is in fact also an infinite
expansion that converges to the expression

\begin{equation}
    W = 
        \left(1-v_{c}\cdot P\right)^{-1}\cdot v_{c}.
\end{equation}
Thus, the action of the screened Coulomb interaction on the pairs of
overlapping MLWFs, defined by the square bracket term in
Eqn. (\ref{eq:wannier W}), may now be cast as an inversion problem
of the form
\begin{equation}
    \int \left(1-A\left(\mathbf{r},\mathbf{r}^{\prime}\right)\right)
    \tau_{vv^{\prime}}\left(\mathbf{r^{\prime}}\right)\,
    \mathrm{d}\mathbf{r}^{\prime} =
        \tau_{vv^{\prime}}^{b}\left(\mathbf{r}\right),
    \label{eq: tau_bvv}
\end{equation}
where $A=v_{c}\cdot P$ and $\tau_{vv^{\prime}}^{b}(\mathbf{r})$ is
the bare Coulomb operator acting on the MLWF pairs: 
\begin{equation}
    \tau_{vv^{\prime}}^{b}\left(\mathbf{r}\right) = 
        \int v_{c}(\mathbf{r},\mathbf{r}^{\prime})
        w_{v}\left(\mathbf{r}^{\prime}\right)w_{v^{\prime}}\left(\mathbf{r}^{\prime}\right)\,
        \mathrm{d}\mathbf{r}^{\prime}.
    \label{vww}
\end{equation}
Finally, $\tau_{vv^{\prime}}(\mathbf{r})$ is the resulting vector describing
the screened Coulomb interaction arising from the Wannier products.
In this way, it is possible to evaluate 
\begin{equation}
    \left|\left\{ 
    \tilde{\xi}^{\prime\prime\prime}
    \right\} \right\rangle = 
        \hat{K}^{d}\left|\left\{ \tilde{\xi}\right\} \right\rangle 
\end{equation}
as
\begin{equation}
    \tilde{\xi}_{v}^{\prime\prime\prime}(\mathbf{r}) = 
        -\int P_{c}(\mathbf{r},\mathbf{r}^{\prime})
        \sum_{v^{\prime}}\tilde{\xi}_{v^{\prime}}(\mathbf{r}^{\prime})
        \tau_{vv^{\prime}}(\mathbf{r}^{\prime})\mathrm{d}\mathbf{r}^{\prime}.
    \label{eq:evaluate Kd}
\end{equation}
In practice, equation \eqref{eq: tau_bvv} is solved iteratively for $\tau_{vv^\prime}$ via a conjugate-gradient minimization. 
At each iteration, the polarizability operator $P$ acts on a generic wavefunction $\tau_{vv^\prime}$, for which established DFPT techniques have been employed; these perform the multiplication exactly, all the while avoiding explicit sums over empty states.~\cite{baroni_phonons_2001, umariPRB2009, umariPRB2010, roccaJCP2010}.

It is worth stressing that this approach bypasses the explicit calculation (and storage on the disk) of the $W$ matrix. In addition, through equation \eqref{eq: s threshold}, we only compute the relevant terms of $W$, and as a result of the inversion procedure these are ensured to be fully converged.

\subsection{Validation}
\label{sec: validation}
To validate our direct minimization strategy for the calculation of the screened Coulomb interaction we compare its performance against a standard \textsc{Gwl} BSE calculation. 
To this end we compute the first $\pi\rightarrow\pi^{*}$ singlet transition energies of an isolated benzene ($\mathrm{C_{6}H_{6}}$) molecule.
The two methods are initialized from the same \textsc{Quantum ESPRESSO} DFT calculation~\footnote{
We use a $20\times20\times20$ \AA\ simulation cell with an total energy cutoff of 70 Ry for plane waves. 
For the atomic cores, we use optimized norm-conserving Vanderbilt pseudopotentials and treat the exchange and correlation using the PBE gradient corrected functional} 
and use the same set of quasiparticle energies (obtained from DFT+KI). 
In addition, for the BSE calculations the threshold $s$ of equation \eqref{eq: s threshold} determines the total number of overlapping MLWFs products;\citep{marsiliPRB2017} in the cases of small molecules such as benzene, which has 15 valence electrons, no spatial cutoff is required and all of the 225 Wannier products are computed.

In the \textsc{Gwl} approach, the $W$ is written in an optimal auxiliary basis set that is given by the eigenvectors of the model static polarizability.\cite{umariPRB2009, umariPRB2010}
This optimal polarizability basis is a truncated sum, and therefore for absolute agreement with the direct approach developed here a complete sum over the eigenvectors would be required.
For a screened interaction, converged to within 0.1 eV, we need 700 eigenvectors, the resulting singlet transition energy is 5.09 eV.
Instead, our direct procedure yields a computed transition energy of 5.02 eV.

\subsection{Scaling in KI-BSE simulations}

In this section, we compare the final scaling of our approach with that of a standard -- plane wave (PW) based -- implementation of the $G_0W_0$-BSE method for charge neutral excitations.
To illustrate this point, we consider a hypothetical system comprising of $N_\mathrm{at}$ atoms in a simulation cell of volume $\Omega_\mathrm{V}$.
The electronic structure, and excited states of this system can be described by $N_v$ valence- and $N_c$ conduction-states expanded in a basis of plane waves with $N_\mathrm{G}$ terms. 
Each of these components of the system increase linearly with the size of the system, in the sense that doubling the number of atoms doubles the number of electrons and therefore $N_v$ and $N_c$ states (below a fixed energetic cutoff).

The most computationally intensive part of the $G_0W_0$ calculation is the evaluation of the polarizability.
Employing the PW based example described above, the construction of the polarizability matrix requires $N^2_{\mathrm{G}} \times N_c \times N_v$ operations and scales as the fourth power of the system size $\mathcal{O}(N^4)$.
Recently, lower scalability $\mathcal{O}(N^3)$ has been reported for the $GW$-step, resting on either non-PW basis sets \cite{foerster_o_2011, wilhelm_toward_2018} or real space-imaginary time techniques \cite{kaltak_low_2014, liu_cubic_2016}.
In the BSE step, the calculation of the excitonic Hamiltonian also involves the multiplication of an order $N^2_\mathrm{G}$ matrix by $N_v\times N_c$-vectors. 
In a standard PW approach, this step scales as $\mathcal{O}(N^4)$ with system size, or as $\mathcal{O}(N^3)$ when utilizing the locality of MLWFs.\cite{marsiliPRB2017}

In our KC+BSE the only parameter is the dimension of the basis set. 
This make the approach considerably easier to manage and suitable for large scale simulations such as material screening and high-throughput simulations in general.
For small systems, where the computational cost is dominated by the FFT, the KI calculation and any orbital-density-dependent calculation in general, scales as $N \times N_\mathrm{DFT}$, where $N_\mathrm{DFT}$ is the scaling of the underlying DFT calculation (typically $N^3$). 
This is essentially because one has to compute $N$ local potential instead of a single KS potential. 
In spite of the fact that this scaling is the same as for a $GW$ calculation, it's important to stress that only the static dielectric matrix is needed 
for the evaluation of the screening coefficients. 
Moreover, in a $GW$ calculation there are typically several convergence parameters linked to the calculation of the frequency dependent dielectric matrix. Instead, since we utilize techniques based on DFPT to overcome explicit summations over empty states, the only parameter in the KI calculation is the dimension of the basis set (the number of plane-waves in our implementation).

The BSE step in the KI-BSE procedure is comprised of two parts, the calculation of $\tau_{vv^\prime}$ and the application of the excitonic Hamiltonian.
To calculate $\tau_{vv^\prime}$ $N^3$ operations for the application of the DFTB machinery are required $N_v \times N_v$ times, scaling overall as $\mathcal{O}(N^5)$ with system size.
Since our method makes use of the locality of MLWFs through equation 17,\cite{marsiliPRB2017} the scaling associated with the calculation of $\tau_{vv^\prime}$ can reduced by one order so that the overall scaling of this step is $\mathcal{O}(N^4)$.
This is not surprising since it corresponds to the $\mathcal{O}(N^4)$ scaling of the $GW$ step we avoid in the calculation of $W$. 
The remaining part, the application of the excitonic Hamiltonian, scales as $\mathcal{O}(N^4)$ or with MLWFs as $\mathcal{O}(N^3)$.\cite{marsiliPRB2017}\\

\section{Results and Discussion}

\subsection{Benchmarking: Thiel's Set}

To test the present approach we calculate the low-energy singlet excitations for each of the molecules in Thiel's set (TS).\citep{schreiberJCP2008,silva-juniorJCP2008} TS is comprised of 28 small organic molecules and limited to just four elements: C, H, O and N\citep{schreiberJCP2008,silva-juniorJCP2008} (see Figure 1 of Ref. \citenum{schreiberJCP2008}). The set is subdivided into four groups based on the functionality (or chemical family) of the molecules; these are the ($i$) unsaturated aliphatic hydrocarbons, ($ii$) aromatic hydrocarbons and heterocycles, ($iii$) aldehydes, keytones and amines and ($iv$) the nucleobases. In recent years TS has emerged as a standard molecular test set for evaluating the calculation of neutral excitations; as such reference data is available for correlated wavefunction methods,\citep{schreiberJCP2008, silva-juniorJCTC2010, silva-juniorMP2010} TD-DFT\citep{silva-juniorJCP2008, jacqueminJCTC2009} and many-body perturbation theory based on $GW\text{+BSE}$\citep{brunevalJCP2015,jacqueminJCTC2015}. 

Rather than comparing with experimental values, it has become standard practice to benchmark to a set of ``best theoretical estimates'' (BTE) for molecular excitations. The original BTEs comprise 104 singlet and 63 triplet excitation energies, each of which take into account results from coupled-cluster theories (CC2, CCSD, CC3), complete-active-space second-order perturbation theory (CAS-PT2) and a degree of human intuition~\cite{schreiberJCP2008}.

Following an investigation into basis set effects on the computed transition energies, the BTEs were subsequently updated  leading to the definition of the reference data: BTE-2~\cite{silva-juniorJCTC2010}. Diffuse transitions were found to be poorly described by the TZVP basis set employed in the generation of the BTEs; the larger $aug$-cc-pVTZ basis set was found to be more accurate in calculating those transitions, and adopted for the BTE-2. Of the transitions computed in the present work, there is a 0.07 eV mean absolute difference between the BTE and BTE-2 reference sets, suggesting that whilst the choice of atomic basis set may be of particular importance for specific molecules, it has a rather small impact on the overall statistical analysis. Consequently, for our analysis we have adopted the BTE-2 energies.

To be consistent with previous work, we carried out our simulations using fixed (MP2-optimized) atomic coordinates provided in the Supporting Information of Ref~\citenum{schreiberJCP2008}. 
We used cubic cells with a lattice dimension of 20 \AA; however, for convergence of the quasiparticle energy levels, some of the molecules required larger simulation cells, specifically  naphthalene (23 \AA), pyrrole (28 \AA) and imidazole (25 \AA).
In the generation of the Kohn-Sham eigenstates, we employed optimized norm-conserving Vanderbilt (ONCV) pseudopotentials~\cite{hamann_optimized_2013} as developed by Schlipf and Gygi~\cite{ schlipf_optimization_2015, ONCV_website} to describe the electron-ion interactions, and the Perdew-Burke-Ernzerhof exchange-correlation functional~\cite{perdew_generalized_1996}. The total energy cutoff for the planewave basis set was determined on a molecule-by-molecule basis ensuring a convergence of 10 meV in the DFT eigenvalues, and sampling reciprocal space at the $\Gamma$-point only.

Owing to the presence of delocalized states in the conduction manifold, it is convenient to make use of a scissor operator in the BSE calculations. 
This rigidly shifts the Kohn-Sham eigenvalues, preserving the energetic ordering of the states making it easier to distinguish relevant transitions from those involving vacuum states.
The size of the scissor operator is physically motivated, according to the magnitude of the KI-corrected energy gap between localized HOMO and LUMO states. 
Unless explicitly stated otherwise, the size of the scissor operators applied for each molecule can be obtained from Table S1 of the  Supporting Information. 
The long-range Coulomb interaction is truncated at a radius equal to half of cell length, and we have not enforced a cutoff threshold for overlapping Wannier pairs, which means that all of the $N_v \times N_v$ products 
expressed in equation \eqref{vww} are computed for each molecule.           

In the following sections we present the results of KI-BSE applied to TS and compare its performance to more common approaches, namely different flavors of $GW$-BSE and TD-DFT.
Beforehand, we provide some considerations that are relevant for a judicious comparison.
In this work, ($i$) the underlying basis set is built from plane waves and, as far
as we know, no previous benchmarking is available utilizing such a basis; 
($ii$) we do not employ any degree of self-consistency in the determination
of the quasiparticle levels, whilst for this set of molecules eigenvalue self-consistency
plays a role in improving agreement with the BTE-2\cite{jacqueminJCTC2015}; 
($iii$) the wavefunctions used to compute the screened Coulomb interaction are taken from the PBE exchange-correlation functional (it has been shown that hybrid exchange-correlation functionals, which typically yield more localised orbitals, provide a better agreement with the BTE~\cite{brunevalJCP2015}).

\begin{table}[t]
\begin{tabular}{l l c c c c c}
\hline
 & & MAE & MAPE & MSE & $r$ & ref.\\
 & & (eV) & (\%) & (eV) & &\\
 \hline
 \multicolumn{2}{l}{KI-BSE:} & \\
 PBE & Plane Waves & 0.54 & 10.6 & -0.43 & 0.94 & Present work\\
 \\
 \multicolumn{2}{l}{$G_0W_0$-BSE:} & \\
 PBE     & TZVP                 & 0.83 & 16.1 & -0.83 & 0.97 & [\citenum{brunevalJCP2015}]   \\
 B3LYP   & TZVP                 & 0.46 & 8.90 & -0.41 & 0.97 & [\citenum{brunevalJCP2015}]   \\
 BHLYP   & TZVP                 & 0.22 & 4.02 &  0.03 & 0.97 & [\citenum{brunevalJCP2015}]   \\
 PBE0    & $aug$-cc-pVTZ$^a$    & 0.61 & 11.5 & -0.59 & 0.98 & [\citenum{jacqueminJCTC2015}] \\
 \\
 \multicolumn{2}{l}{ev$GW$-BSE:} & \\
 PBE     & $aug$-cc-pVTZ$^a$   & 0.28 & 5.16 & -0.18 & 0.97 & [\citenum{jacqueminJCTC2015}] \\
 PBE0    & $aug$-cc-pVTZ$^a$   & 0.27 & 4.91 & -0.15 & 0.97 & [\citenum{jacqueminJCTC2015}] \\
\\
 \multicolumn{2}{l}{TD-DFT:} & \\
 PBE     & TZVP                & 0.55 & 10.7 & -0.47 & 0.95 & [\citenum{jacqueminJCTC2009}] \\
 B3LYP   & TZVP                & 0.27 & 5.03 & -0.09 & 0.97 & [\citenum{jacqueminJCTC2009}] \\
 PBE0    & TZVP                & 0.26 & 4.72 &  0.04 & 0.97 & [\citenum{jacqueminJCTC2009}] \\
 BHLYP   & TZVP                & 0.48 & 9.03 &  0.41 & 0.94 & [\citenum{jacqueminJCTC2009}] \\
 PBE0    & $aug$-cc-pVTZ$^a$   & 0.23 & 4.41 & -0.08 & 0.97 & [\citenum{jacqueminJCTC2015}] \\ 
 \hline 
\end{tabular}
\caption{Statistical analysis summarizing a selection of reference data available for TS. The mean averaged error (MAE), mean averaged percentage error (MAPE), mean signed error (MSE) and correlation coefficient $r$ are reported. Each approach is compared with the set BTE-2, where, in order to make direct comparisons with our results, each data set is truncated to the 89 transitions obtained in the KI-BSE study.\\ $^a$Amide molecules were computed using the cc-pVTZ basis in this data.}
\label{ref_stats}
\end{table}

A direct comparison between KI-BSE and $G_0W_0$-BSE computed at the PBE level with the same underlying basis set would be ideal; however, to the best of our knowledge no previous benchmark that makes use of PWs is available. 
A summary of some of the reported benchmarks for TS is reported in Table \ref{ref_stats}. 
The work of Bruneval \emph{et al.}, which shows how increasing amounts of exact-exchange in the underlying DFT calculation can be used to improve the performance of the computed BSE transition energies (see Table \ref{ref_stats}), reports $G_0W_0$-BSE results at the PBE level using the TZVP basis set\cite{brunevalJCP2015}. 
As discussed, the TZVP basis set underperforms for excited states with diffuse orbitals (convergence testing shows BSE transition energies can be overestimated by up to 0.65 eV with respect to larger atomic bases~\cite{brunevalJCP2015}). 
However, despite such large deviations, we demonstrated in the case of the quantum chemical reference data that the statistical analysis of the entire set is negligibly affected when moving from TZVP to larger atomic bases.   
Alternatively, Jacquemin \emph{et al.} also make use of PBE wavefunctions in their ev$GW$-BSE simulations, with the advantage that basis set errors are ameliorated through the use of $aug$-cc-pVTZ (and cc-pVTZ) for all molecules.\cite{jacqueminJCTC2015} Whilst this improves the transition energies, in contrast with the present perturbative KI-approach, Jacquemin \emph{et al.} impose an eigenvalue self-consistency cycle in the calculation of the Green's function in the $GW$ step of their calculations. This additional self-consistency step results in an updated description of the screening and potentially, markedly different quasiparticle energy levels.
Therefore, in order to maintain maximum consistency between approaches, we take the energies computed by Bruneval \emph{et al} as $G_0W_0$-BSE reference data at the PBE level; these values are reproduced in Section 2 of the Supporting Information. 

On the other hand, many TD-DFT benchmarks for TS are available in the literature, including an extensive survey of 29 DFT exchange-correlation functionals\cite{jacqueminJCTC2009}. In Table \ref{ref_stats} we summarise the performance of several popular functionals including PBE. In spite of the fact that a TD-PBE (TZVP) benchmark has been reported, given the significant differences between MBPT and TD-DFT approaches we instead adopt the best performing TD-PBE0 ($aug$-cc-pVTZ) set as a second reference as this will allow us to analyze our results in the context of the current state-of-the-art.

\subsubsection{Quasiparticle levels and energy gaps}

\begin{table}[t]
\begin{tabular}{l l r r}
				&			& PBE 	& KI \\ \hline
Thiel's set		& MAE (eV) 	&  3.30 & 0.19 \\
				& MAPE (\%)	& 35.96 & 2.14 \\ \hline
\emph{Series 1} & MAE (eV) 	&  3.20	& 0.22 \\
				& MAPE (\%)	& 35.96	& 2.67 \\
\emph{Series 2} & MAE (eV)	&  3.17 & 0.11 \\
				& MAPE (\%)	& 35.28 & 1.25 \\
\emph{Series 3} & MAE (eV)	&  4.06	& 0.28 \\
				& MAPE (\%)	& 40.21	& 2.79 \\
\emph{Series 4} & MAE (eV)	&  2.84	& 0.25 \\
				& MAPE (\%)	& 32.49	& 2.87 \\ \hline

\end{tabular}
\caption{The mean absolute errors (MAE, eV) and mean absolute percentage errors (MAPE) for the computed vertical ionization potential reported with respect to experimental values. NB: Propanamide has been omitted from the analysis as no experimental VIP was reported.}
\label{koopmans_mae}
\end{table}

Turning first to the quality of the quasiparticle energies obtained using the KC functional, we note that in an earlier investigation of the G2-1 molecular test set the KI approach was found to yield vertical ionization potentials (VIPs) with a mean absolute error of approximately 0.5 eV with respect to experiment.\cite{borghiPRB2014} In the present work, for TS, the mean absolute error (MAE), which is reported in Table \ref{koopmans_mae}, is instead less than 0.2 eV corresponding to a mean absolute percentage error (MAPE) of the order of 2\%. The greater accuracy found for the KI functional with respect to experimental VIPs in this work is due to a better description of screening via orbital-density dependent coefficients $\{\alpha_i\}$~\cite{colonna_screening_2018}, and to the restriction of TS to strictly organic compounds with a limited number of atomic species. 

Considering individually each of the different chemical families the KI functional performs consistently well across the entire molecular set with a largest MAPE of 2.87\% for \emph{Series 4} which contains the four nucleobases. In terms of accuracy, this is slightly better than previous $G_0W_0$ results (MAE: 0.34 eV\cite{jacqueminJCTC2009}, 0.29 eV\cite{qianprb2011}; MAPE: 3.95 \%\cite{jacqueminJCTC2009}, 3.33 \%\cite{qianprb2011}) for the same four molecules. 

As far as we are aware, reference data permitting a direct comparison of KI and $GW$ VIPs for all 28 molecules in TS is unavailable. We instead turn to the computed HOMO-LUMO quasiparticle gaps for the $G_0W_0$ data set at the PBE level, which are reported in Table S1 of the Supporting Information. For KI, we report the energy difference between the KI HOMO and LUMO levels, whereas for $G_0W_0$ data we reproduce values reported in the Supporting Information of Reference~\citenum{brunevalJCP2015}. The effect of the quasiparticle corrections is to open the gap with respect to the PBE value. Initial comparison suggests that the two different methods are not completely consistent: across TS the mean absolute difference (MAD) between KI and $G_0W_0$ is fairly large (0.43 eV), with absolute differences ranging from 0.02 eV ($E$-Butadiene) to 1.67 eV (Acetamide). Looking in more detail, we find that in general the computed gaps are comparable and only four of the molecules exceed a 1 eV difference; these are pyrrole, imidazole, acetamide and propanamide. Excluding these molecules from the statistical analysis would yield better agreement with the MAD = 0.27 eV. 

Closer inspection of the simulations for pyrrole, imidazole, acetamide and propanamide reveals that at the PBE level the LUMOs in these molecules are spatially delocalized, which results in unusually small quasiparticle corrections at the KI level.
On the other hand, when a small localized basis set is used (for instance, the one used in the $G_0W_0$ results we are comparing with), these delocalized states may be completely missing due to the limited capability of the basis set in reproducing very diffuse orbitals. 
In fact, if the delocalized LUMO is disregarded, the resulting quasiparticle gaps, reported in Table S1 of the supporting information, are in agreement with the $G_0W_0$ reference energies, suggesting that for these four molecules the LUMO state in a finite basis set actually corresponds to the LUMO+1 in a plane-wave approach.
Therefore, for these molecules it is necessary to treat the localized LUMO+1 explicitly in the calculation of the KI corrections and to derive a KI scissor operator for the higher states based on the KI energy of the LUMO+1 (and not on the KI energy of the LUMO).

\subsubsection{Singlet transition energies}

\begin{figure*}[t]
\includegraphics[width=\textwidth]{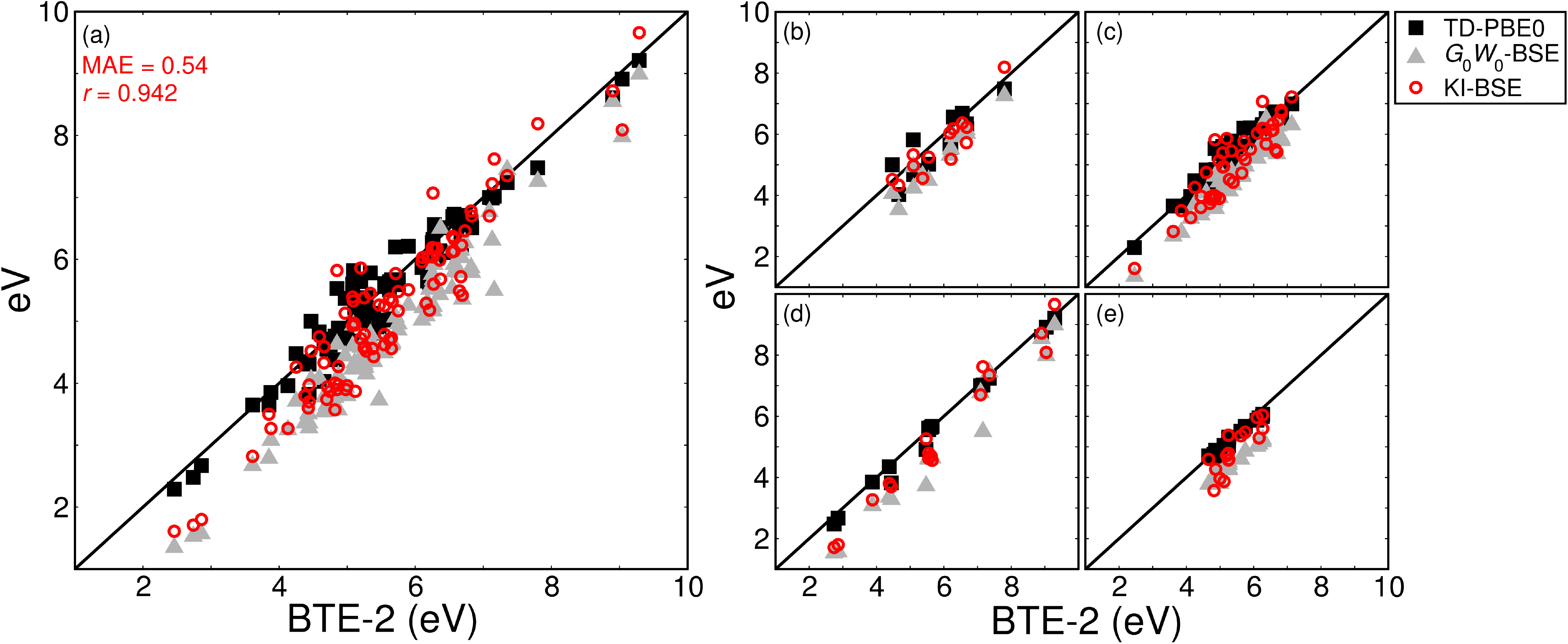}
\caption{Plots of our computed KI-BSE transition energies (red circles) as a function of the BTE-2\cite{} reference data. Previous benchmarking of TD-PBE0\cite{jacqueminJCTC2009} (black squares) and $G_0W_0$-BSE \cite{brunevalJCP2015} (grey triangles) methods are also plotted. In (a) the full set of 89 transition energies are plotted, in (b-e) they are resolved according to chemical family: unsaturated aliphatic hydrocarbons; aromatic hydrocarbons \& heterocycles; aldehydes, keytones \& amides and nucleobases respectively. Inset the KI-BSE MAE (eV) and correlation coefficient $r$ are reported.}
\label{vs_bte2}
\end{figure*}

Through a post-processing analysis of the ground-state occupied, virtual and excitonic wavefucntions we have been able to identify (based on orbital symmetries) 89 of the 104 singlet excitation energies reported in the BTE-2 data set. This represents 85.6 \% of the energies; similar to the work by Jacquemin \emph{et al}\cite{jacqueminJCTC2015}, our analysis failed to identify the remaining transitions in the BSE spectrum with a high degree of confidence. Notable differences in our analysis and that of Jacquemin \emph{et al} are the inclusion of the naphthalene $A_g$ transition (BTE-2: 6.49 eV) and the omission of the acetone $B_1$ (BTE-2: 9.04 eV) and acetamide $A^\prime$ (BTE-2: 7.14 eV) transitions. The transition energies are presented in Figure \ref{vs_bte2}(a), where we plot the correlation between the KI-BSE, $G_0W_0$-BSE and TD-PBE0 energies and the BTE-2 energies.

The KI-BSE scheme has a MAE (0.54 eV) and a mean signed error MSE (-0.43 eV) that, when viewed in the context of the results in Figure \ref{vs_bte2}a, points to a systematic underestimation of the transition energies with respect to the best theoretical estimates. Yet, the large correlation coefficient (0.940) is an indication that the approach accurately reproduces trends in the transition energies. 

In comparison with other schemes, KI-BSE outperforms $G_0W_0$-BSE at the PBE level in terms of both the MAE (-0.84 eV) and MSE (0.84 eV). In fact, inspection of Figure \ref{vs_bte2}a reveals that in almost all cases KI-BSE energies are closer to the BTE-2 than the $G_0W_0$-BSE are. Whilst we acknowledge the limitations associated with the performance of the TZVP basis set in the description of diffuse states, looking at Table \ref{ref_stats}, in particular at TD-PBE0 results with different atomic bases ($\Delta\text{MAE} = 0.03$ eV) and the 0.07 eV MAD between BTE and BTE-2, it can be seen that this has a limited impact on the overall statistics for all of TS. Therefore, it can be asserted that, not only does the method introduced here remove the need for extensive and time-consuming convergence testing, but the KI-BSE significantly ($\approx 0.3$ eV) improves on the popular $G_0W_0$-BSE approach in the case of molecular systems.  Rather more surprisingly, when compared with other $G_0W_0$-BSE benchmarks, the KI-BSE data also outperforms simulations at the level of PBE0 (MAE = 0.61 eV; MSE = -0.59 eV)\cite{jacqueminJCTC2015} in spite of a previous observation that fractions of exact exchange within the DFT exchange-correlation functional ($\text{PBE}0=25 \%$) have been linked to improved BSE energies\cite{brunevalJCP2015}. It is worth noting that the quality of the KI-BSE data is comparable with results for $G_0W_0$-BSE at the level of B3LYP (20 \% exact exchange), which has a reported MAE $\approx$ 0.5 eV.

On the other hand, Figure \ref{vs_bte2} and the computed MAE (0.23 eV) and MSE (-0.08 eV) suggest that KI-BSE underperforms when compared with the TD-PBE0~\cite{jacqueminJCTC2015} data set. 
Where KI-BSE is systematically underestimating the transition energies the TD-PBE0 values tend to bracket the BTE-2 data as demonstrated in Figure \ref{ref_stats} and evidenced by the much lower MSE (-0.08 eV). 
From a purely statistical standpoint, the KI-BSE and TD-PBE~\cite{jacqueminJCTC2009} approaches perform almost identically, although we note that marginal improvements to the TD-PBE dataset may potentially be obtained through the use of a different basis set. 
We also anticipate that the present method could be improved by combining other Koopmans-compliant functionals with the BSE.
As an example, the KIPZ functional, which is built on top a Perdew-Zunger self-interaction corrected functional~\cite{perdew_self-interaction_1981}, has been found to improve on the KI quasiparticle energies MAPEs~\cite{borghiPRB2014,nguyenPRL2015, colonna_koopmans-compliant_2019}.
It is reasonable to expect that the combination of an improved description of the quasiparticle eigenspectrum and updated Koopmans-compliant eigenstates will lead a better description of molecular excitonic spectrum (this will be the topic of a future investigation).

\subsection{Optical absorption spectrum of $\text{C}_{60}$}

We have further validated the direct calculation of the charge screening and benchmarked the KI-BSE approach for the computation the optical absorption spectra.
In the \textsc{Gwl} code, whilst individual excitonic states are obtained from a conjugate-gradient minimization, the complete complex part of the dielectric matrix is iteratively computed following the Lanczos method~\cite{ankudinov_parallel_2002,roccaJCP2010}. 
Owing to previous benchmarking of our $GW$-BSE implementation\cite{marsiliPRB2017}, the C$_{60}$ molecule is a convenient and readily available choice of a larger molecular system for testing these two novel aspects. In this section we compare optical spectra computed using explicit $W$ calculations and the direct screening procedure introduced in Section \ref{sec: screening}. We then compare spectra computed with the $G_0W_0$-BSE and KI-BSE methods.

Simulations are carried out in a cubic cell with a lattice dimension of 40 Bohr, the atomic coordinates are taken from previous work~\cite{qianPRB2015} and in the DFT calculations, core levels are represented by ONCV pseudopotentials~\cite{hamann_optimized_2013, schlipf_optimization_2015, ONCV_website} and Kohn-Sham states are generated using the PBE exchange correlation functional\cite{perdew_generalized_1996}.
No spatial truncation has been applied to the summation entering into the evaluation of the $K^d$ term. Moreover, in the $GW$-BSE calculations the static polarizability operator is constructed from a basis of 2000 elements\cite{umariPRB2009, umariPRB2010}.
For the evaluation of the performance of our direct screening method, we use a scissor operator corresponding to a 4.94 eV gap, which is based on the iterative eigenvalue self-consistent $GW$ calculations in reference \citenum{qianPRB2015}. Instead, in the comparison of KI-BSE with $G_0W_0$-BSE, the respective scissor operators are taken from the KI and $G_0W_0$~\cite{qianPRB2015} quasiparticle gaps.

\begin{figure}[t]
\includegraphics[]{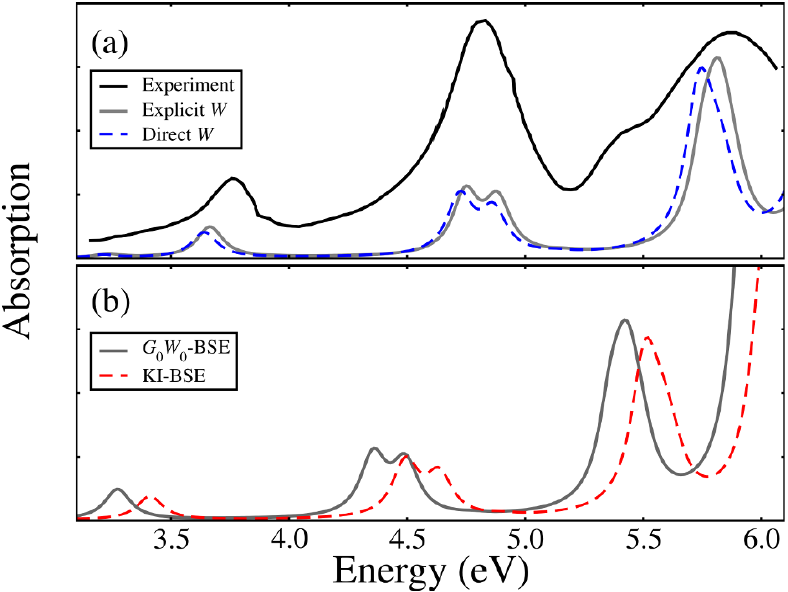}
\caption{Plots of the absorption spectra for the $\text{C}_{60}$ molecule: 
(a) A comparison between spectra obtained from experimental measurement (black, solid)~\cite{bauernschmittJACS1998}, previous $G_0W_0$-BSE simulations (grey, solid)~\cite{marsiliPRB2017} and our KI-BSE implementation (blue, dashed). The theoretical spectra have been computed using the same scissor operator taken from~\cite{qianPRB2015}. (b) A comparison of the spectra computed from $G_0W_0$-BSE (grey, solid)~\cite{marsiliPRB2017} and KI-BSE (blue, dashed) using a scissor operator derived from the respective $G_0W_0$~\cite{qianPRB2015} and KI gap.}
\label{c60spectrum}
\end{figure}

Figure \ref{c60spectrum}a plots the comparison between the experimental optical spectra, and the spectra computed with the explicit (grey) and direct (blue) screening. 
We find excellent agreement in the peak positions, with the largest deviation between theoretical results still within 0.05 eV for the peaks located at approximately 5.75 eV in energy. 
This result is not surprising, the differences between the two spectra are in line with difference in the energies of the first excited states of the benzene molecule computed in Section \ref{sec: validation}. 
Instead, the spectra plotted in Figure \ref{c60spectrum}b compare results of KI-BSE (red) and $G_0W_0$-BSE (grey) simulations. 
Both are significantly red shifted compared to experimental absorption spectra\cite{bauernschmittJACS1998} seemingly suggesting that self-consistency steps could be important at the quasipartice level in the BSE procedure. 
However, it should be noted that the peaks in the KI-BSE spectrum appear at higher energies than the analogous $G_0W_0$-BSE ones, this is due to the slightly larger KI quasiparticle gap, which reduces the underestimation of the peak energies with respect to Figure \ref{c60spectrum}a. 

\section{Conclusions}

To summarize, this work introduces an alternative route towards the computation of charge-neutral excitations within many-body perturbation theory that avoids the use of the $GW$ approximation.
The method couples Koopmans-compliant orbital-dependent density functionals, which are used to compute quasiparticle energies, with the Bethe-Salpeter equation. 
This approach can be used to compute excitonic eigenstates directly via iterative minimization techniques such as steepest-descent or conjugate-gradients; alternatively absorption spectra can be calculated using the Lanczos-chain method~\cite{ankudinov_parallel_2002}. 
For the treatment of metallic or small-gap solids, the methodology presented could also be expanded to include both a dynamical description of the screened Coulomb interaction\cite{umari_infrared_2018} and considering the full (non-Tamm-Dancoff approximation) excitonic Hamiltonian.\cite{roccaJCP2010} 

In comparison with standard $GW$-BSE implementations, the present approach offers several advantages: KI-BSE $(i)$ makes use of techniques from DFPT in order to avoid explicit summations over the empty state manifold~\cite{baroni_phonons_2001,roccaJCP2010}, $(ii)$ avoids the need for $GW$ (and lengthy convergence testing), $(iii)$ does not require the computation and storage of the zero-frequency $W$ matrix and $(iv)$ directly computes the fully converged screened Coulomb interaction only for the relevant terms of the excitonic Hamiltonian.

KI-BSE performs better for Thiel's test set than comparable $G_0W_0$-BSE calculations using the same underlying DFT functional and performs well when compared with state-of-the-art hybrid TD-DFT simulations. Further improvements may be achieved by using non-perturbative Koopmans-compliant functionals, such as the KIPZ functional, which accounts for relaxations in the eigenstates used in the solution of the BSE. Finally, the method can be combined with the recent implementation of the KI (and other KC) functionals for extended systems to study charge-neutral excitations in the solid-state~\cite{nguyen_koopmans-compliant_2018}. 

\section{Acknowledgements}

J.D.E and P.U. acknowledge funding from the EU-H2020 research and 
innovation programme under grant agreement No 654360 NFFA-Europe. 
N.C. and N.M. acknowledge the Swiss National Science Foundation 
(SNSF), through Grant No. 200021-179139, and its National Centre
of Competence in Research (NCCR) MARVEL. 
We acknowledge PRACE for awarding us access to Marconi at CINECA, Italy.

\section{Supporting Information}

Tables containing complete sets of DFT energies, KI quasiparticle energies and Bethe-Salpeter Equation transition energies are available in the supporting information.

\bibliography{main}

\end{document}